\documentclass[aps,pra,twocolumn,superscriptaddress,english,showpacs]{revtex4-1}
\usepackage{graphicx}
\usepackage{amssymb}
\usepackage{graphics}
\usepackage{times}
\usepackage{amstext}
\usepackage{bbm}
\usepackage{babel}

\begin{document}

\title{Quantum Defragmentation Algorithm}

\author{Daniel Burgarth}
\affiliation{Institute for Mathematical Sciences, Imperial College London, SW7 2PG, United Kingdom
\\ QOLS, The Blackett Laboratory, Imperial College London, Prince Consort Road, SW7 2BW, United Kingdom
}
\author{Vittorio Giovannetti}
\affiliation {NEST,  Scuola Normale Superiore \& CNR-INFM, Piazza
		dei Cavalieri 7, I-56126 Pisa, Italy}

\begin{abstract}
In this addendum of our paper [D. Burgarth and V. Giovannetti, Phys. Rev.
Lett. \textbf{99}, 100501 (2007)] we prove that during the transformation that allows one to
enforce  control by relaxation on a quantum system, the ancillary memory  can be kept at a finite
size, independently from the fidelity one wants to achieve. The result is obtained by introducing the quantum analog of 
defragmentation algorithms which are employed for efficiently reorganizing classical information in conventional hard-disks.
Our result also implies that the reduced dynamics in any noisy system can be simulated with finitely many resources.  
\end{abstract}

\maketitle 

Accomplishing controllability of quantum mechanical systems is one of
the main hurdles towards a large scale quantum computer. 
In recent years,  increasing attention has been devoted in developing schemes
that allows on to achieve global control on a large many-body quantum system $V= C \cup \overline{C}$ by 
only having direct access to a relatively small subpart $C$ of it~\cite{Lloyd2004,Romano2006,kay2008,Schirmer2008,BGB2007,BBBG,Burgarth2007,Burgarth2007c,CONT,CONT2,KAY}.
In this contest, the majority of  results obtained so far have been derived within the general framework of 
``algebraic"   approach to control theory, e.g. see Ref.~\cite{BOOK}. Here the allowed operations are parametrized by specifying which (local) components of the system Hamiltonian 
can be manipulated via proper choices of classical control pulses. 

%%%%%%%%%%%%%%%%%%%%%%%%%%%%%%%%%%%%%%%%%
\begin{figure}
\includegraphics[width=8cm]{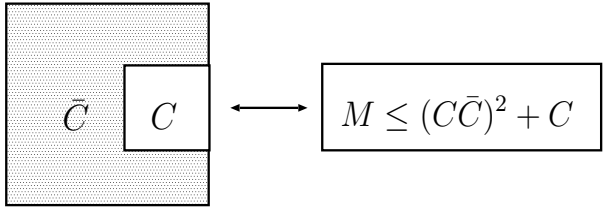}
\caption{\label{fig:schemes}
Schematics of the control by relaxation scheme. The control on the large system $V= C \cup \overline{C}$ is exerted through an auxiliary (fully controllable) quantum memory $M$ 
which is directly coupled to the subsystem $C$.}
\end{figure}
%%%%%%%%%%%%%%%%%%%%%%%%%%%%%%%%%%%%%%%%%

An independent approach was recently proposed by us  in Refs.~\cite{Burgarth2007,Burgarth2007c}, introducing the notion of 
 local controllability of quantum
systems via ``relaxation". In this scheme
additional control of a large ancillary system $M$ was assumed and a
method of controlling $V= C \cup \overline{C}$  by acting on $C$ \emph{and }$M$
was suggested. This is essentially achieved by transferring the states
of $V$  into $M$ through a sequence of iterative   operational steps (see Fig.~\ref{fig:schemes}) which induces an effective relaxation of $V$ into the memory degree of freedom. The
states are then controlled in $M$ and, using the inverted sequence of steps, transferred back to
$V$.  

On one hand, such new method can be important in inhomogeneous scenarios, where some parts of the system are easier to control than others.  
It also allows for an easy-to-check criterion if a given system is controllable which can be applied analytically to large  systems~\cite{Burgarth2007,Burgarth2007c} and which was subsequently 
generalized to the algebraic control scenario~\cite{BBBG}.
Finally  compared to algebraic control it has the advantage that the control protocol is constructive and follows a clear physical intuition.

On the other hand, the main drawback of  the controllability by relaxation  approach 
stems from the fact that 
it  cannot reduce the size of the
controlled system (in contrast to algebraic control). Indeed 
 to be able to store arbitrary states  from $V$ to the ancillary memory $M$ the latter must be at least as large as the former (i.e. $\mbox{dim}M\geq \mbox{dim}V$).
Even more problematic is the fact that up to now no  upper bounds were known on the minimal size of $M$ which is needed to accomplish the
control.  In this paper we fix this problem by 
showing that $M$  can be kept at a finite
size, which is maximally twice as large as $V$. This
is a major improvement to~\cite{Burgarth2007,Burgarth2007c}, where $M$ was
arbitrarily large. 
The result is derived by introducing the quantum analogous  of 
defragmentation algorithms. 
In computer science, defragmentation is a process that allows one to reduce the 
amount of fragmentation in file systems. This is obtained by reorganizing the contents of the disk to store the pieces of each file close together and contiguously
while creating larger regions of free space.
Here  we use a  similar idea to (coherently) 
 compress quantum information in the quantum memory $M$ during 
 its transferring from $V$. This  results in  
  a more efficient storing of messages, which saves valuable memory space for the subsequent 
 data processing transformations.

\section{The Algorithm} \label{sec:algo}

Whilst referring to
Refs.~\cite{Burgarth2007,Burgarth2007c} for details, the scheme of
control by relaxation can be summarized by saying that it consists in
a {\em downloading} stage in which $C$ is iteratively
coupled to a fixed, finite-dimensional subspace (say a qubit) $M_1$ of
$M$ that is re-prepared into a fiduciary state $|0\rangle_{M_1}$ after
each iteration.  The $\ell$-th step of this process is described by a
unitary downloading operation $W_{\ell}$, which for large $\ell$ moves arbitrary
states $|\psi\rangle_{C\overline{C}}$ of the system into the memory,
i.e.,
\begin{eqnarray}
W_{\ell}|\psi\rangle_{C\overline{C}}\otimes|0\rangle_{M}
\approx|0\rangle_{C\overline{C}}\otimes|\Phi(\psi)\rangle_{M}, 
\label{winf}
\end{eqnarray}
with $|\Phi(\psi)\rangle_M$ being a linear function of the input state
$|\psi\rangle_{C\overline{C}}$. 
 They are then controlled in $M$ and
moved back to the system in an {\em uploading stage} that reverses the process~(\ref{winf}).
It is worth stressing that the transformations $W_\ell$ are known and are independent from the input state of the system. 

The above introduction seems to indicate that indeed
$M=C\overline{C}$ is large enough to contain images of all possible
states. This is not the case as states are only transferred asymptotically and for intermediate $\ell$, the downloading operator
$W_{\ell}$ is generating entanglement between $C\overline{C}$ and
$M$. However introducing an orthonormal basis 
$\{|k\rangle_{C\overline{C}} \}$ of $C\overline{C}$, a generic state $|\psi\rangle=\sum_k\alpha_k |k\rangle_{C\overline{C}}$ 
after $\ell$ steps can be written as
\begin{eqnarray}
W_{\ell}\sum_{k}\alpha_{k}|k\rangle_{C\overline{C}}\otimes|0\rangle_{M}=
\sum_{kk'}\alpha_{k}\omega_{kk'}^{(\ell)}|k' 
\rangle_{C\overline{C}}\otimes|\xi_{kk'}^{(\ell)}\rangle_{M},
 \nonumber \\
\label{defw}
\end{eqnarray}
with $|\xi_{kk'}^{(\ell)}\rangle_M$ being a set of $\left(\dim
C\overline{C}\right)^{2}$ not-necessarily orthogonal vectors of $M$.
Independently of the value of $\ell$, the states $\{
|\xi_{kk'}^{(\ell)}\rangle_M\}_{kk'}$ span a space of dimension
smaller than or equal to $\left(\dim C\overline{C}\right)^{2}$: they
can thus be fitted into a subsystem $M_0$ of $M$ which is twice as
large as $C\overline{C}$. Therefore, by including an extra 
{\em defragmentation} step into the protocol of
Ref.~\cite{Burgarth2007}, the memory can be kept at a finite size. In
detail, write $M = M_0 \otimes M_1$. The defragmentation consists
then in operating on the memory with a unitary transformation  which
maps the $|\xi_{kk'}^{(\ell)}\rangle_M$ into states of the form
$|\tilde{\xi}_{kk'}^{(\ell)}\rangle_{M_0} \otimes |0\rangle_{M_1}$
with $|0\rangle_{M_1}$ being the fiduciary state of the downloading
stage, while the $|\tilde{\xi}_{kk'}^{(\ell)}\rangle_{M_0}$ are instead
characterized by having the same mutual scalar product as the
$|\xi_{kk'}^{(\ell)}\rangle_M$, i.e. 
\begin{eqnarray}
{_{M_0}\langle}  \tilde{\xi}_{k''k'''}^{(\ell)}   |\tilde{\xi}_{kk'}^{(\ell)}\rangle_{M_0} = 
{_{M} \langle} \xi_{k''k'''}^{(\ell)} |\xi_{kk'}^{(\ell)}\rangle_M
 \;,
 \end{eqnarray}
for all $k,k',k''$, and $k'''$.
The whole procedure can be
iterated easily by observing that at the $(\ell+1)$-th step the state of
the system can be still described as in Eq.~(\ref{defw}) for a proper
choice of the vectors $|\xi_{kk'}^{(\ell+1)}\rangle_M$.

It is worth noticing that the defragmentation procedure presented
here finds also useful application in the context of spin chain
communication~\cite{Bose2007}.  Indeed by generalizing the result of
the end-gate protocol of Ref.~\cite{BGB2007,ZHANG} to the
multi-excitation sector case, it shows that the memory-assisted
transmission scheme of Ref.~\cite{GB2006} can be implemented with
finite resources.

\begin{acknowledgments}
We acknowledge fruitful discussions with C. Bruder and S. Bose. This work is supported by  
%T. Schulte-Herbr\"uggen, support by the QIP-IRC and Wolfson College Oxford
%(D.B.), the EPSRC, UK, the Royal Society and the Wolfson
%Foundation (S.B.),  the EC IST-FET project EuroSQUIP, the Swiss
%NSF, and the NCCR Nanoscience (C.B.), and the Quantum Information
%research program of Centro di Ricerca Matematica Ennio De Giorgi of
%Scuola Normale Superiore (V.G.).
the Italian Ministry of University and Research under the FIRB-IDEAS 
project RBID08B3FM (VG) and by the EPSRC grant EP/F043678/1 (DB).
\end{acknowledgments}


\begin{thebibliography}{10}

\bibitem{Lloyd2004} S. Lloyd, A.~J. Landahl, and J.-J.~E. Slotine,
Phys. Rev. A \textbf{69}, 012305 (2004).



\bibitem{Romano2006}R. Romano and D. D' Alessandro, 
Phys. Rev. A {\bf 73}, 022323 (2006).

\bibitem{kay2008} A. Kay, Phys. Rev. A \textbf{78}, 012346 (2008). 

\bibitem{Schirmer2008}S. G. Schirmer, I. C. H. Pullen, and P. J. Pemberton-Ross,
Eprint arXiv:0801.0721 [quant-ph].


\bibitem{BGB2007} D. Burgarth, V. Giovannetti, and S. Bose, 
Phys. Rev. A {\bf 75}, 062327 (2007).
 
 \bibitem{BBBG}D. Burgarth, S. Bose, C. Bruder, and V. Giovannetti, Phys. Rev. A {\bf 79}, 06030(R) (2009).


\bibitem{Burgarth2007} D. Burgarth and V. Giovannetti, Phys. Rev.
Lett. \textbf{99}, 100501 (2007).

\bibitem{Burgarth2007c} D. Burgarth and V. Giovannetti,  in
{\em Quantum Information and Many Body Quantum Systems}, proceedings, M. Ericsson and S. Montangero (eds.), Pisa, Edizioni della Normale, p. 17 (2008); 
Eprint arXiv:0710.0302 [quant-ph].

\bibitem{KAY} A. Kay, Phys.Rev. Lett. {\bf 98}, 010501(2007). 

\bibitem{CONT}D. Burgarth, K. Maruyama, M. Murphy, S. Montangero, T. Calarco, F. Nori, M. B. Plenio, Phys. Rev. A {\bf 81} 040303(R) (2010.

\bibitem{CONT2}A. Kay and P. J. Pemberton-Ross, Phys. Rev. A \textbf{81}, 010301(R) (2010).



\bibitem{BOOK} D. D¡ÇAlessandro, {\em Introduction to Quantum Control and Dynamics} (Taylor and Francis, Boca Raton, 2008). 


%
%\bibitem{kane1998} B. E. Kane, Nature {\bf 393}, 133 (1998).

%
%\bibitem{Albertini2002} F. Albertini and D. D'Alessandro, Linear
%Algebra and its Applications \textbf{350}, 213 (2002).

%\bibitem{Ramakrishna95} V. Ramakrishna, M. V. Salapaka, M. Dahleh,
%H. Rabitz, and A. Pierce, Phys. Rev. A \textbf{51}, 960 (1995).

%\bibitem{Schirmer2001} S.~G. Schirmer, H. Fu, and A.~I. Solomon,
%Phys. Rev. A \textbf{63}, 063410 (2001);
%S.~G. Schirmer, I.~C.~H. Pullen, and A.~I.
%Solomon, J. Phys. A {\bf 35}, 2327 (2002).

%\bibitem{Schulte}T. Schulte-Herbr\"{u}ggen, A. Sp\"{o}rl, N. Khaneja, 
%and S. J. Glaser, Phys. Rev. A \textbf{72}, 042331 (2005).

%\bibitem{Fitzsimons}J. Fitzsimons, and J. Twamley, Phys. Rev. Lett.
%\textbf{97}, 090502 (2006).

%\bibitem{Benjamin} S. C. Benjamin and S. Bose, Phys. Rev. Lett. \textbf{90},
%247901 (2003); S. C. Benjamin and S. Bose, Phys. Rev. A \textbf{70},
%032314 (2004).

%

%\bibitem{Raussendorf}R. Raussendorf, Phys. Rev. A \textbf{72}, 052301
%(2005).

%

%

%

%
%\bibitem{Schol2005} U. Schollw\"{o}ck, Rev. Mod. Phys. {\bf 77}, 259 (2005); F. Verstraete, V. Murg, and J. I. Cirac, Adv. Phys. {\bf 57}, 
%143 (2008); J. I. Cirac and F. Vestraete, arXiv:0910.1130 [cond-mat.str-el]. 


  
 \bibitem{Bose2007} S. Bose, Contemporary Phys. {\bf 48}, 13 (2007).



%\bibitem{optical} E. Jan\'{e}, G. Vidal, W. D\"ur, P. Zoller, and
%  J. I. Cirac, Quantum Inf. Comput. {\bf 3}, 15 (2003); L.-M. Duan,
%  E. Demler, and M. D. Lukin, Phys. Rev.  Lett. {\bf 91}, 090402
%  (2003).
%
%\bibitem{arrays} M. J. Hartmann, F. G. S. L. Brand\~ao, and M. B. Plenio, 
%Nature Physics {\bf 2}, 849 (2006). 
%
%\bibitem{arraysb}D.G. Angelakis, M.F. Santos, and S. Bose, Phys. Rev. A (R) 76, 031805 (2007). 

%\bibitem{solid} A. Romito, R. Fazio, and C. Bruder, Phys. Rev. B {\bf 71}, 
%100501(R) (2005). 

%\bibitem{Aazami2008} A. Aazami, {\em Approximation algorithms and hardness
%  for the propagation problem} (unpublished).

%\bibitem{Severini2008} S. Severini, Eprint arXiv:0805.0181 [quant-ph].

%\bibitem{AKLT} Y. Xu and V. E Korepin. Eprint arXiv:0805.3542 [quant-ph].

%\bibitem{SU3} A. Bayat and V. Karimipour, Phys. Rev. A 75, 022321 (2007).



\bibitem{ZHANG} J. Zhang, N. Rajendran, X. Peng, and D. Suter,
  Phys. Rev. A {\bf 76}, 012317 (2007).


\bibitem{GB2006} V. Giovannetti and D. Burgarth, Phys. Rev. Lett. {\bf 96}, 
 030501 (2006).





\end{thebibliography}
\end{document}